\documentclass[conference]{IEEEtran}
\IEEEoverridecommandlockouts

\usepackage{cite}
\usepackage{amsmath,amssymb,amsfonts}
\usepackage{algorithmic}
\usepackage{graphicx}
\usepackage{textcomp}
\usepackage{xcolor}
\usepackage{xurl}
\usepackage{amsfonts}
\usepackage{siunitx}
\usepackage{booktabs}
\usepackage{multirow}
\usepackage{colortbl}
\usepackage{amssymb}
\usepackage{pifont}
\newcommand{\cmark}{\ding{51}}
\newcommand{\xmark}{\ding{55}}
\usepackage{graphicx}
\usepackage{array}
\usepackage{graphicx}
\usepackage{makecell}

\def\BibTeX{{\rm B\kern-.05em{\sc i\kern-.025em b}\kern-.08em
    T\kern-.1667em\lower.7ex\hbox{E}\kern-.125emX}}

\begin{document}
\topmargin=0mm 
\title{CARD: Cross-component Audio Representation Distillation for Encoder-Free Audio Captioning \\
}

\author{
\IEEEauthorblockN{Ganesh Pavan Kartikeya Bharadwaj Kolluri\textsuperscript{1},
Yuchen Zhang\textsuperscript{1,2},
Michael Kampouridis\textsuperscript{1},
Ravi Shekhar\textsuperscript{1,2}}
\IEEEauthorblockA{
\textsuperscript{1}\textit{School of Computer Science and Electronic Engineering, University of Essex} \\
\textsuperscript{2}\textit{Institute for Analytics and Data Science, University of Essex} \\
\{karthik.kolluri, yuchen.zhang, mkampo, r.shekhar\}@essex.ac.uk}
}

\maketitle

\begin{abstract} 
Modern automated audio captioning systems pair a frozen audio encoder with a large language model (LLM) via a trainable projector, incurring the encoder's inference cost and
bottlenecking the model through its fixed acoustic features. We present CARD, an encoder-free audio captioning model that removes the encoder at inference: a 13.2M projector feeds a frozen LLM with merged LoRA adapters, while the teacher used to train it is
discarded. CARD distills a pre-trained audio teacher (CLAP-HTSAT) into the model, but rather than injecting it into the LLM alone, it routes the teacher's representations across components: perceptual stages to the projector and semantic stages to the LLM. This placement improves CIDEr-D by +11.9 over an LLM-only distilled model on AudioCaps and by +5.0 on Clotho, reaching 55.4 against a 66.4 encoder-kept upper bound with no encoder at
inference, showing that where a teacher's knowledge is placed matters as much as its presence. 
\end{abstract}


\section{Introduction}

Automated audio captioning (AAC) generates a natural-language description of the sound events in an audio clip, such as ``rain falls as thunder rumbles in the distance.'' Unlike automatic speech recognition (ASR), which transcribes
spoken words, AAC aims to describe the broader acoustic scene, capturing non-speech sounds and their relationships in free-form natural language. This ability to summarize what a clip contains has made AAC valuable for accessibility, audio retrieval and content understanding~\cite{Kim2019AudioCapsGC, Drossos2019ClothoAA}.

Most recent AAC systems approach this task with one architectural design. A frozen pre-trained audio encoder such as CLAP~\cite{Elizalde2023CLAPLA, wu2023laionclap} converts the raw clip into acoustic feature vectors, and a trainable projector maps them into the input space of a Large Language Model (LLM) that decodes them into a caption. Because both pre-trained components stay frozen, the projector is the only component trained, and it learns just the mapping from encoder outputs into the language model's input space, which is why the design has been so widely adopted~\cite{Kim2024EnCLAPCN, Chen2024SLAMAACEA, Li2024DRCapDC}.
Keeping the encoder, however, means carrying it at inference. Every clip must complete a full forward pass through a second pre-trained model before the language model can act, and the language model perceives audio only through the features that encoder preserves.

Discarding the encoder removes that dependency and frees the language model from the encoder's fixed features, but it also removes the foundation the projector and the language model build on, leaving them to perceive audio directly from the raw acoustic signal. Learning to extract acoustic features from scratch while also learning to generate captions is unstable, and encoder-free models have consequently struggled to match their encoder-based counterparts~\cite{Diao2024EVE, Wang2025VisionAL}. Knowledge distillation offers a way to keep the encoder's knowledge without keeping the encoder itself: a pre-trained model supervises the student during training and is then discarded, so the student inherits the teacher's knowledge without carrying it at inference~\cite{Wang2025VisionAL}. Existing approaches, however, direct that supervision almost entirely at the language model~\cite{Cheng2026AuRA}, leaving the projector to learn from the captioning objective alone even though it produces every audio token the language model consumes. This raises a question prior work has not examined: how teacher knowledge should be distributed across the components of an encoder-free AAC model.

We address this with CARD, a cross-component distillation framework that assigns teacher representations to student components according to their functional roles. Earlier teacher layers carry lower-level acoustic cues and
supervise the audio projector, while later layers carry semantic abstraction and supervise the language model, so each component is taught at the level it is equipped to learn. Once training ends, the teacher and all distillation heads are removed and the LoRA adapters are merged into the base model, leaving only the projector and the LLM, with no audio encoder.

Our contributions are as follows:
\begin{itemize}
\item We propose \textbf{CARD}, an encoder-free audio captioning model deployed as a 13.2M projector and a LoRA-merged Qwen3-4B, shrinking the learned audio interface from 34.4M to 13.2M parameters.\footnote{Code is available at \url{https://github.com/KarthikKolluriKB/CARD}.}
\item We introduce \textbf{cross-component distillation}, which routes the teacher's early, perceptual stages to the projector and its later, semantic stages to the language model, and verify with linear probes and centered kernel alignment (CKA) that each component receives the stages it can absorb.
\item We show that \textbf{placement is decisive}: CARD gains +11.9 CIDEr-D on AudioCaps and +5.0 on Clotho over LLM-only distillation, which itself gains only +0.3 and +0.2 over no distillation, while routing the full teacher hierarchy to the projector is worse than not distilling. CARD reaches 55.4 against a 66.4 encoder-kept upper bound and transfers zero-shot to Clotho-AQA~\cite{Lipping2022ClothoAQAAC}.
\end{itemize}

\section{Related Work}

\textbf{Encoder-based AAC.}
Recent LLM-based AAC systems share a common design in which a frozen pre-trained audio encoder extracts acoustic representations and a trainable projector aligns them with a language decoder: EnCLAP~\cite{Kim2024EnCLAPCN} pairs a frozen CLAP encoder with a BART decoder, DRCap~\cite{Li2024DRCapDC} adds retrieval-augmented prompting, SLAM-AAC~\cite{Chen2024SLAMAACEA} couples the EAT encoder to an LLM through a lightweight projector, and LOAE~\cite{Liu2024EnhancingAA} uses the CED~\cite{Dinkel2023CEDCE} encoder. These systems caption strongly, but keeping a separate encoder carries three costs. It remains active at inference, adding computation and memory beyond the language model itself~\cite{Fan2026MelLLM}; it confines the language model to its frozen representation space, so acoustic information the encoder discards is inaccessible downstream~\cite{adept2023fuyu, Wang2025VisionAL}; and bridging the two requires a projector trained on paired audio--text data to align encoder outputs with the LLM embedding space~\cite{Kim2024EnCLAPCN, Chen2024SLAMAACEA}. These costs motivate asking whether the encoder can be removed entirely.

\textbf{Encoder-free models and distillation.}
Encoder-free multimodal LLMs internalize modality understanding within the language model itself, eliminating dedicated encoders at inference. The paradigm was first explored in vision, where Fuyu-8B~\cite{adept2023fuyu}, EVE~\cite{Diao2024EVE} and Mono-InternVL~\cite{Luo2024MonoInternVL} learn visual representations inside an LLM from lightweight projections of raw images, and VoRA~\cite{Wang2025VisionAL} adds teacher-guided distillation, supervising LoRA adapters in the LLM with a frozen vision transformer through block-wise hidden-state alignment before merging them and discarding the teacher. Knowledge distillation~\cite{Hinton2015DistillingTK}, which transfers teacher knowledge through output- or representation-level supervision, thus becomes the mechanism by which an encoder-free architecture is achievable rather than a compression step~\cite{Gu2023MiniLLMOD}. In audio, Mel-LLM~\cite{Fan2026MelLLM} and Gemma~4~(12B)~\cite{gemma4_2026} rely on downstream supervision alone: the former projects Mel-spectrogram patches through a linear projector into a LoRA-adapted LLM and learns acoustic representations from speech tasks such as ASR and Text-to-Speech (TTS), the latter projects raw audio directly into the LLM at larger scale, and neither supervises the audio frontend. AuRA~\cite{Cheng2026AuRA}, concurrent with our work, adopts VoRA's~\cite{Wang2025VisionAL} strategy for speech, distilling a frozen ASR encoder into LoRA-adapted language layers through layer-wise hidden-state alignment, but its supervision is applied entirely on the language-model side, and it targets transcription rather than the diverse environmental and non-speech events AAC must describe. In both vision and audio, then, the teacher supervises only the language model; the projector that turns raw signals into tokens is left to learn on its own. Whether each component should instead receive supervision suited to its role remains an open question, and it is the one we take up here.

\begin{figure*}[!t]
\centering

\includegraphics[width=\textwidth,height=0.31\textheight]{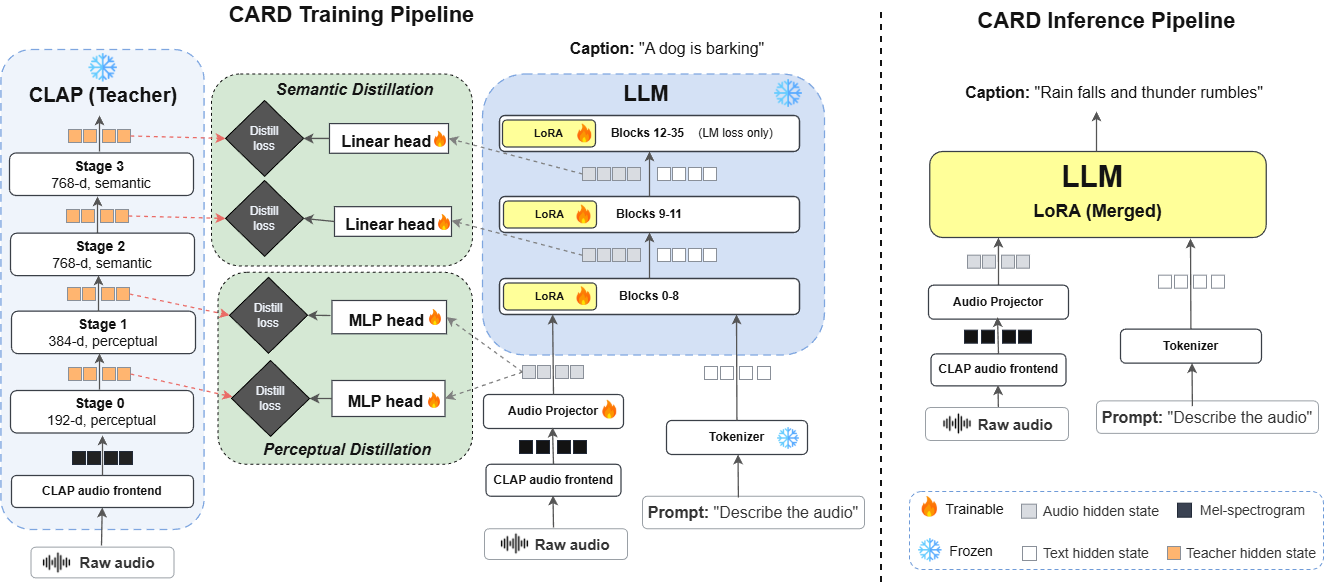}

\caption{Overview of CARD. During training (left), a frozen CLAP teacher supervises the two student components by role: its early, perceptual stages (0-1) distill into the audio projector, and its later, semantic stages (2-3) into the LLM's LoRA adapters. At inference (right), the teacher and distillation heads are dropped, and the LoRA is merged into the LLM, leaving only the projector and the LLM, with no audio encoder. }
\label{fig:arch}
\vspace{-0.5em}
\end{figure*}

\section{Methodology} 

CARD is an audio captioning model that eliminates the need for a dedicated audio encoder. Unlike conventional architectures~\cite{Chen2024SLAMAACEA} that rely on a pre-trained audio encoder, projector, and LLM decoder, CARD learns acoustic knowledge through representation-level distillation from a frozen teacher model. The student model consists only of an audio projector and a LoRA-adapted LLM. After training, the teacher is removed, and the LoRA adapters are merged into the LLM, resulting in an efficient architecture comprising only the LLM backbone and audio projector while maintaining audio-captioning capability. The overview of  CARD is shown in Figure~\ref{fig:arch}.

\subsection{Model Architecture}
\label{sec:arch}

\textbf{Student model.}
The student model consists of a lightweight audio projector followed by a LoRA-adapted large language model. Given an input audio clip, the projector converts the audio into a sequence of audio tokens, which are prepended to the text prompt embeddings and consumed by the language model to generate the output caption. The projector is a stack of strided one-dimensional convolutions over the log-Mel spectrogram, producing a sequence of audio tokens at the language model's hidden size; the full configuration is given in Section~\ref{sec:setup}.

We adopt Qwen3-4B~\cite{qwen3} as the language backbone. Adaptation is performed using LoRA~\cite{hu2022lora} on the linear layers of every transformer block, including the query, key, value, output, gate, up, and down projections. All other language-model parameters are frozen; the audio projector and the distillation heads are trained jointly with the LoRA adapters. After training, the LoRA weights are merged into the backbone, introducing no additional parameters or computational overhead at inference.

\textbf{Teacher model.}
We employ a frozen CLAP~\cite{wu2023laionclap} as the teacher throughout training. CLAP is contrastively pre-trained on paired audio--text data and serves as the audio encoder in several state-of-the-art AAC systems~\cite{Kim2024EnCLAPCN,Li2024DRCapDC}. Its representations therefore provide caption-relevant acoustic knowledge for supervising the student model.

The CLAP audio branch adopts an HTSAT backbone~\cite{chen2022htsat}, implemented as a hierarchical Swin Transformer~\cite{liu2021swin}. The backbone produces a hierarchy of intermediate representations with progressively increasing semantic abstraction. During training, the teacher receives the same log-Mel spectrogram as the student. The resulting intermediate representations are used as supervision signals in the component-aware distillation described next.

\subsection{Cross-Component Distillation}
\label{sec:xdistill}

In existing teacher-guided encoder-free audio models~\cite{Cheng2026AuRA}, teacher supervision is typically applied only to language-model layers, leaving the audio projector trained solely through the downstream captioning objective. Since every audio token consumed by the language model is produced by the projector, insufficient supervision at the projector may limit the quality of all downstream representations. 

CARD addresses this mismatch by aligning the teacher hierarchy with the functional structure of the student. The CLAP teacher provides acoustic representations at multiple levels of abstraction through its hierarchical HTSAT backbone, while the student separates computation across two components: the projector transforms raw acoustic signals into language-compatible representations, and the language model interprets these representations for semantic reasoning and caption generation. Teacher supervision is therefore assigned according to component function: we extract the outputs of the four HTSAT stages, $\{t_0,t_1,t_2,t_3\}$, and use the earlier, lower-level acoustic representations to supervise the projector and the later, higher-level semantic ones to supervise the language model. Section~\ref{sec:analysis} gives the probe and CKA measurements that justify this split.

For projector supervision, we attach one two-layer MLP head per stage ($2560\!\rightarrow\!768\!\rightarrow\!d_i$, GELU) to the projector's audio tokens $u_{\mathrm{proj}}$, mean-pooling each head's output over time:
\begin{equation*}
\hat{u}_0 = h_0(u_{\mathrm{proj}}), \quad
\hat{u}_1 = h_1(u_{\mathrm{proj}}),
\end{equation*}
where $h_i$ denotes the head for stage $i$. The results are matched to the mean-pooled teacher features $t_0$ and $t_1$ by cosine distillation.

For language-model supervision, we use the hidden representations at the audio-token positions from the lower 12 transformer blocks. Let $u_{\mathrm{llm}}^{1}$ denote the average representation over blocks 0--8, and let $u_{\mathrm{llm}}^{2}$ denote the average representation over blocks 9--11. Each language-model head is an RMSNorm followed by a single linear projection to the corresponding teacher stage dimension:
\begin{equation*}
\hat{u}_2 = W_2 u_{\mathrm{llm}}^{1}, \quad
\hat{u}_3 = W_3 u_{\mathrm{llm}}^{2}.
\end{equation*}
Similarly, the projected representations $\hat{u}_2$ and $\hat{u}_3$ are compared with the corresponding teacher representations $t_2$ and $t_3$ using the cosine distillation loss. The remaining transformer blocks are not directly supervised by the teacher and are optimized only through the captioning objective.

The pairwise distillation loss for each teacher--student representation pair is defined as
\begin{equation*}
\ell_{\mathrm{distill}}(\hat{u}_i,t_i)
=
1-\cos(\hat{u}_i,t_i),
\end{equation*}

where $i\in\{0,1,2,3\}$ indexes the four teacher stages, and $\hat{u}_i$ and $t_i$ denote the projected student representation and the corresponding teacher representation. For the projector stages these are single pooled vectors, while for the language-model stages the teacher is interpolated along time to the audio-token length and the cosine is averaged over token positions.

\subsection{Training Objective}
\label{sec:objective}

CARD is trained in two consecutive phases with different optimization objectives. During Phase~1, the student model is jointly optimized with a captioning loss, $\mathcal{L}_{\mathrm{cap}}$, a projector distillation loss, $\mathcal{L}_{\mathrm{proj}}$, and a language-model distillation loss, $\mathcal{L}_{\mathrm{llm}}$, while the CLAP teacher remains frozen. The audio projector, LoRA adapters, and distillation heads are jointly updated to transfer caption-relevant acoustic knowledge from the teacher into the encoder-free student. After Phase~1, the teacher model and all distillation heads are removed.

Specifically, the projector distillation loss is obtained by averaging the pairwise distillation losses associated with the lower teacher stages,

\begin{equation*}
\mathcal{L}_{\mathrm{proj}}=\frac{1}{2}\left(
\ell_{\mathrm{distill}}(\hat{u}_0,t_0)+\ell_{\mathrm{distill}}(\hat{u}_1,t_1)\right).
\end{equation*}

The language-model distillation loss is computed from the higher teacher stages,
\begin{equation*}
\mathcal{L}_{\mathrm{llm}}
=
\frac{1}{2}
\left(
\ell_{\mathrm{distill}}(\hat{u}_2,t_2)
+
\ell_{\mathrm{distill}}(\hat{u}_3,t_3)
\right).
\end{equation*}

The captioning loss is the standard autoregressive cross-entropy loss over the target caption tokens,
\begin{equation*}
\mathcal{L}_{\mathrm{cap}}
=
-\sum_{i=1}^{N}
\log p(y_i \mid y_{<i},x),
\end{equation*}
where $x$ denotes the input audio tokens and $y=\{y_1,\ldots,y_N\}$ is the target caption.

The overall training objective during Phase~1 is
\begin{equation*}
\begin{aligned}
\mathcal{L}_{phase1}
=
\mathcal{L}_{\mathrm{cap}}
+
\lambda_{\mathrm{proj}}
\mathcal{L}_{\mathrm{proj}}
+
\lambda_{\mathrm{llm}}
\mathcal{L}_{\mathrm{llm}},
\end{aligned}
\end{equation*}

where $\lambda_{\mathrm{proj}}=\lambda_{\mathrm{llm}}=1$ in all experiments. The teacher model and all distillation heads are used only during training and are discarded afterwards. At inference, only the audio projector and the LoRA-merged language model are retained.

During Phase~2, the student model is fine-tuned separately for each evaluation dataset using only the captioning loss $\mathcal{L}_{\mathrm{cap}}$, allowing the model to adapt to the target caption distribution without additional teacher supervision:
$\mathcal{L}_{phase2}=\mathcal{L}_{\mathrm{cap}}$.

After Phase~2, the LoRA adapters are merged into the language-model backbone, yielding the final deployed model with only the lightweight audio projector and the language model.

\section{Experiments}

\subsection{Experimental Setting}

\textbf{Datasets.}
Table~\ref{tab:data} summarizes the datasets used in the two training phases. Phase~1 uses a large-scale mixture of captioned-audio datasets for representation learning, including WavCaps~\cite{mei2024wavcaps}, Auto-ACD~\cite{sun2024autoacd},
AudioCaps~\cite{Kim2019AudioCapsGC}, Clotho~\cite{Drossos2019ClothoAA}, and
MACS~\cite{martin2021macs}. To preserve the language understanding capability of the backbone LLM, text-only instruction data from Infinity-Instruct~\cite{li2025infinityinstruct} are interleaved during training.  
Phase~2 fine-tunes the pre-trained student separately for each evaluation benchmark, following the two-phase setup of SLAM-AAC~\cite{Chen2024SLAMAACEA}: for AudioCaps evaluation, the model is fine-tuned on the prompt-expanded AudioCaps training set~\cite{Kim2019AudioCapsGC}, while Clotho evaluation uses the corresponding prompt-expanded Clotho training set~\cite{Drossos2019ClothoAA}. We evaluate on the standard AudioCaps and Clotho benchmarks.

\begin{table}[t]
\centering
\caption{Dataset summary.}
\label{tab:data}
\small
\setlength{\tabcolsep}{4pt}
\renewcommand{\arraystretch}{0.85}
\begin{tabular}{l l r}
\toprule
\textbf{Phase} & \textbf{Dataset} & \textbf{Samples} \\
\midrule
\multirow{6}{*}{Phase 1}
& WavCaps~\cite{mei2024wavcaps}       & 140{,}750 \\
& Auto-ACD~\cite{sun2024autoacd}      & 82{,}268 \\
& AudioCaps~\cite{Kim2019AudioCapsGC} & 45{,}178 \\
& Clotho~\cite{Drossos2019ClothoAA}   & 3{,}839 \\
& MACS~\cite{martin2021macs}          & 3{,}930 \\
& Text-instruction mix                & 50{,}000 \\
\midrule
\multirow{2}{*}{Phase 2}
& AudioCaps                            & 225{,}890 \\
& Clotho                               & 95{,}975 \\
& Clotho-AQA~\cite{Lipping2022ClothoAQAAC}                           & 21{,}114 \\
\midrule
\multirow{2}{*}{Evaluation}
& AudioCaps test                       & 975 \\
& Clotho evaluation                    & 1{,}045 \\
& Clotho-AQA test              & 1{,}442 \\
\bottomrule
\end{tabular}
\vspace{-0.5em}
\end{table}

\textbf{Implementation details.}
\label{sec:setup}
CARD uses Qwen3-4B~\cite{qwen3} as the language backbone and CLAP~\cite{wu2023laionclap} as the Phase~1 teacher. LoRA is applied to all seven linear projections of all 36 transformer blocks of Qwen3-4B, with rank $r=16$ and scaling factor $\alpha=16$. Audio clips are resampled to 48\,kHz and cropped or repeat-padded to 10\,s, from which the CLAP processor produces a 64-band log-Mel spectrogram ($n_{\mathrm{fft}}$ 1024, hop 480, 50--14000\,Hz). The projector turns this into audio tokens with two stride-2 convolutions (kernel 3, $64\!\rightarrow\!512\!\rightarrow\!2560$, GELU), followed by LayerNorm, a $2560\!\rightarrow\!2560$ linear layer, and learned positional embeddings, for 13.2\,M parameters and 4$\times$ temporal downsampling. On the distillation side, the projector heads are two-layer MLPs with hidden width 768 and the language-model heads are single linear projections, mapping to teacher stage dimensions of 192 and 384 for the projector and 768 for both language-model stages.

Phase~1 runs for one epoch. 
Optimization is performed using AdamW with $\beta=(0.9,0.95)$, an effective batch size of 64, an initial learning rate of $2\times10^{-4}$, a constant-with-warmup learning-rate schedule (200 warmup steps), and gradient clipping with a maximum norm of $1.0$. The five audio datasets are sampled in proportion to their size, and text-only instruction examples are interleaved as modality-pure batches, never mixed with audio within a batch. Dropout is zero throughout, including in the LoRA adapters and distillation heads. 

In Phase~2, the Phase~1 LoRA update is first merged into the base weights, fixing the learned alignment in the backbone. 
A fresh LoRA adapter ($r{=}16$, $\alpha{=}16$) is then trained on this consolidated backbone for one epoch at a learning rate of $2\times10^{-5}$ with a cosine schedule (50 warmup steps), while the projector continues from its Phase~1 checkpoint.

At evaluation, captions are generated using beam search with a beam size of 4 and a maximum length of 40 tokens. All experiments are conducted on 2×A6000 GPUs with 48 GB memory each.

\begin{table*}[!th]
\centering
\caption{Audio captioning performance (\%) on AudioCaps and Clotho Datasets. 
}

\label{tab:main}
\setlength{\tabcolsep}{4.5pt}
\renewcommand{\arraystretch}{1.25}
\small

\begin{tabular}{l l ccc cccc cccc}
\toprule

\multirow{2}{*}{\textbf{Model}} &
\multirow{2}{*}{\textbf{Variant}} &
\multirow{2}{*}{\textbf{Enc.}}&
\multirow{2}{*}{\textbf{LoRA}}&
\multirow{2}{*}{\textbf{Distill}}&
\multicolumn{4}{c}{\textbf{AudioCaps}} &
\multicolumn{4}{c}{\textbf{Clotho}} \\
\cmidrule(lr){6-9}
\cmidrule(lr){10-13}
 &
 &
 &
 &
 &
 CD & SD & SC & MT & CD & SD & SC & MT \\
\midrule

\multirow{2}{*}{{SLAM-AAC}}

& No LoRA &
\cmark &
\xmark &
- &
59.8 & 38.7 & 17.5 & 23.7 &
32.0 & 22.3 & 12.2 & 16.4 \\

& LoRA &
\cmark &
\cmark &
- &
66.4 & 41.9 & 17.3 & 23.3 &
39.0 & 25.8 & 12.7 & 16.2 \\

\midrule

\multirow{5}{*}{CARD}

& \textit{No Distill } &
\xmark &
\cmark &
No teacher &
43.2 & 27.9 & 12.6 & 18.4 &
22.3 & 15.5 & 8.6 & 13.0 \\

& \textit{LLM Distill} &
\xmark &
\cmark &
LLM only &
43.5 & 27.8 & 12.2 & 18.4 &
22.5 & 15.1 & 8.4 & 13.0 \\

& \textit{Proj\_Full} &
\xmark &
\cmark &
All stages$\to$Projector &
40.7  & 26.4  & 12.1  & 17.8  & 21.2 & 14.6 & 8.1  & 12.7   \\

& \textit{Proj\_Early} &
\xmark &
\cmark &
Early stages$\to$Projector &
52.5 & 33.3 & 14.1 & 20.1 &
24.3 & 16.6 & 8.9 & 13.1 \\

& \textit{Reversed Routing}  &
\xmark & \cmark &
Early$\to$LLM, Later$\to$Proj &
50.0 & 31.8 & 13.5 & 19.3 &
22.8 & 15.9 & 9.0 & 13.2 \\

\cline{2-13}
& \textit{CARD$^\diamond$} &
\xmark &
\cmark &
Projector(all stages) + LLM &
49.9 & 32.0 & 14.1 & 20.3 &
24.8 & 17.1 & 9.4 & 13.1 \\

\rowcolor{gray!15} &
\textbf{\textit{CARD$^*$}} &
\xmark &
\cmark &
Projector(early stages) + LLM &
\textbf{55.4} & \textbf{35.2} & \textbf{15.1} & \textbf{21.2} &
\textbf{27.5} & \textbf{18.8} & \textbf{10.1} & \textbf{14.2} \\

\bottomrule
\end{tabular}
\vspace{-0.5em}
\end{table*}

\textbf{Baselines.}
We use SLAM-AAC as the encoder-based baseline. For a fair comparison, we replace its original audio encoder with the same CLAP encoder used in our teacher model. We report two settings, without LoRA and with LoRA, both of which retain the frozen CLAP encoder during inference.

\textbf{Metrics.}
We use CIDEr-D (CD)~\cite{vedantam2015cider} as the primary metric, as it measures consensus with human references and is widely reported in AAC benchmarks. We also report SPIDEr (SD)~\cite{liu2017spider}, SPICE (SC)~\cite{anderson2016spice}, and METEOR (MT)~\cite{banerjee2005meteor} to provide a comprehensive assessment of caption quality.

\subsection{Overall Results}

Table~\ref{tab:main} compares the encoder-based SLAM-AAC baseline with CARD under several distillation configurations: \emph{No Distill} removes teacher supervision entirely, \emph{LLM Distill} supervises only the language model, \emph{Proj\_Full} and \emph{Proj\_Early} supervise only the audio projector with all four HTSAT stages $\{t_0,t_1,t_2,t_3\}$ or the early stages $t_0,t_1$ respectively, and \emph{CARD}$^{\diamond}$ and \emph{CARD}$^{*}$ add language-model supervision to each, \emph{CARD}$^{*}$ being the configuration described in Section~\ref{sec:xdistill}.

\textbf{Encoder-based baseline.}
The encoder-based SLAM-AAC baseline achieves the strongest overall performance. With LoRA, it reaches 66.4\% CIDEr-D on AudioCaps and 39.0\% on Clotho. This is expected because the CLAP encoder is retained during inference and directly provides acoustic representations to the language model.

\textbf{Full CARD model.}
Combining projector and language-model supervision gives the best encoder-free results. \emph{CARD}$^{\diamond}$ reaches 49.9\% CIDEr-D on AudioCaps, while \emph{CARD}$^{*}$ reaches 55.4\%. On Clotho, \emph{CARD}$^{*}$ also achieves the best encoder-free result, reaching 27.5\% CIDEr-D. Compared with \emph{No Distill}, \emph{CARD}$^{*}$ improves CIDEr-D by 12.2 percentage points on AudioCaps and 5.2 percentage points on Clotho, with consistent gains across SPIDEr, SPICE, and METEOR. To validate these improvements, we conducted significance tests across all comparisons in Table~\ref{tab:main}. \emph{CARD}’s gains over no distillation, LLM-only distillation, and mismatched projector supervision are statistically significant on both datasets, confirming that the results are robust and that the teacher location is critical. 

Overall, the results show that teacher supervision is most effective when early teacher representations supervise the audio projector and later teacher representations supervise the language model. This supports the component-aware distillation strategy used in \emph{CARD}$^{*}$.

\textbf{Effect of distillation placement.}
Removing the encoder drops \emph{No Distill} substantially, to 43.2\% CIDEr-D on AudioCaps and 22.3\% on Clotho. Supervising only the language model recovers almost nothing (\emph{LLM Distill}: 43.5\% and 22.5\%), so language-model supervision alone cannot compensate for the absence of direct supervision on the audio projector. Projector supervision is more effective, but only when the teacher representation matches the projector's role: \emph{Proj\_Full} reaches 40.7\% on AudioCaps while \emph{Proj\_Early} improves this to 52.5\%, and on Clotho \emph{Proj\_Early} improves over \emph{Proj\_Full} by 3.1 CIDEr-D. \emph{Reversed Routing} confirms that direction matters, not merely the presence of projector supervision: it sends the late stages to the projector and the early stages to the language model with data, schedule, LoRA rank and loss weights fixed, so each component still receives two teacher stages, and reaches 50.0\% on AudioCaps and 22.8\% on Clotho at comparable Phase-1 LM loss. 

\textbf{Inference efficiency.} Profiling on a single A6000 (batch 1, beam 4, 50 clips of 10\,s) shows that removing the encoder reduces the learned audio interface from 34.4\,M to 13.2\,M parameters, its latency from 11.1\,ms to 0.4\,ms, and its FLOPs from 12.8\,G to 5.4\,G; the shared log-Mel front-end (39.9\,ms) is parameter-free and unchanged. CARD's 251 audio tokens raise LLM prefill from 550\,G to 2090\,GFLOPs, but decoding dominates, so end-to-end latency (0.62\,s vs 0.63\,s, RTF 0.062) and peak memory (8.4\,GB vs 8.9\,GB) are comparable. Since CARD's audio interface is fixed regardless of teacher size, larger encoders would give proportionally larger reductions.

\subsection{Model Analysis}
\label{sec:analysis}

\textbf{Why the stage assignment works.}
The teacher's stages differ in what they encode, and the projector can only reach some of them. Linear probes on mean-pooled stage features show sound-event predictability rising with depth (mAP 0.243 to 0.422, chance 0.125) while low-level acoustic predictability stays flat, and pairwise CKA between stages decays with separation (0.80 for $t_0$/$t_1$, 0.22 for $t_0$/$t_3$), so the early stages are perceptual and the late stages semantic. CKA between the projector's audio tokens and the teacher stages is 0.68 and 0.74 for $t_0$ and $t_1$ against 0.55 and 0.28 for $t_2$ and $t_3$, an ordering already present at random initialization: the early stages are a target the projector can represent, the late stages are not, which is why routing the full hierarchy to it (\emph{Proj\_Full}) underperforms.

\textbf{Effect of LoRA Capacity.}
Table~\ref{tab:lora_ablation} studies the effect of LoRA capacity under different training settings. With cross-component distillation, the model is sensitive to the LoRA rank. Using rank 8 gives 33.3\% CIDEr-D on AudioCaps and 17.8\% on Clotho, below the \emph{No Distill} setting at rank 16. This suggests that adapter capacity is insufficient for jointly fitting the captioning and distillation objectives. Increasing the rank to 16 gives the best results on both datasets, reaching 55.4\% on AudioCaps and 27.5\% on Clotho. Increasing the rank further to 32 reduces performance to 50.0\% and 24.9\%, showing that larger adapters do not necessarily improve transfer.

This sensitivity is much weaker without cross-component distillation. Under \emph{LLM Distill}, increasing $\alpha$ from 16 to 32 reduces CIDEr-D from 43.5\% to 41.4\% on AudioCaps and from 22.5\% to 20.8\% on Clotho. Without teacher supervision, increasing the LoRA rank from 16 to 32 changes CIDEr-D only slightly, from 43.2\% to 43.9\% on AudioCaps and from 22.3\% to 22.6\% on Clotho. These results suggest that LoRA capacity matters most when the adapter must accommodate both the captioning objective and the cross-component teacher signal.

\begin{table}[!h]
\centering
\caption{Effect of LoRA rank and scaling on \normalfont{CIDEr-D}. 
}

\label{tab:lora_ablation}
\begin{tabular}{l l c c c c}
\toprule
\textbf{Model} & \textbf{Variant} & \textbf{$r$} & \textbf{$\alpha$} & \textbf{AudioCaps} & \textbf{Clotho} \\
\midrule
\multirow{7}{*}{CARD}
& \multirow{2}{*}{\textit{No Distill}}
& 16 & 16 & 43.2 & 22.3 \\
&
& 32 & 32 & 43.9 & 22.6 \\ \cline{2-6}
& \multirow{2}{*}{\textit{LLM Distill}}
& 16 & 16 & 43.5 & 22.5 \\
&
& 16 & 32 & 41.4 & 20.8 \\
\cline{2-6}

& \multirow{3}{*}{\textit{CARD$^{*}$}}
& 8  & 8  & 33.3 & 17.8 \\
&

& \cellcolor{gray!15}16 & \cellcolor{gray!15}16 & \cellcolor{gray!15}\textbf{55.4} & \cellcolor{gray!15}\textbf{27.5} \\

&
& 32 & 32 & 50.0 & 24.9 \\
\bottomrule
\end{tabular}
\vspace{-0.5em}
\end{table}

\begin{table}[!h]
\centering
\caption{Effect of the two training phases at $r=16$.}
\label{tab:stages}
\setlength{\tabcolsep}{5pt}
\renewcommand{\arraystretch}{1.3}
\small
\resizebox{\columnwidth}{!}{
\begin{tabular}{l cccc cccc}
\toprule
\multirow{2}{*}{\textbf{Setting}} & \multicolumn{4}{c}{\textbf{AudioCaps}} & \multicolumn{4}{c}{\textbf{Clotho}} \\
\cmidrule(lr){2-5}\cmidrule(lr){6-9}
 & CD & SD & SC & MT & CD & SD & SC & MT \\
\midrule

Phase 1 Only      & 10.3 & 10.6 & 11.0 & 16.2 & 6.3 & 6.2 & 6.0 & 11.2 \\

Phase 2 Only     & 37.2 & 24.2 & 11.2 & 17.2 & 16.5 & 13.3 & 6.8 & 12.1 \\

\rowcolor{gray!15}
\textbf{Phase 1+2 } & \textbf{55.4} & \textbf{35.2} & \textbf{15.1} & \textbf{21.2} & \textbf{27.5} & \textbf{18.8} & \textbf{10.1} & \textbf{14.2} \\
\bottomrule
\end{tabular}}
\vspace{-0.5em}
\end{table}

\textbf{Effect of Two-Phase Training.}
Table~\ref{tab:stages} reports the contribution of the two training phases. Using only Phase~2 fine-tuning achieves 37.2\% CIDEr-D on AudioCaps and 16.5\% on Clotho. This suggests that caption supervision alone is not sufficient for learning strong audio representations in the encoder-free setting. Using only Phase~1 pretraining performs much worse, reaching 10.3\% on AudioCaps and 6.3\% on Clotho. This indicates that distillation pretraining alone does not adapt the model to the target captioning format.

The full two-phase training pipeline performs best across all reported metrics. It reaches 55.4\% CIDEr-D, 35.2\% SPIDEr, 15.1\% SPICE, and 21.2\% METEOR on AudioCaps. On Clotho, it reaches 27.5\% CIDEr-D, 18.8\% SPIDEr, 10.1\% SPICE, and 14.2\% METEOR. Compared with Phase~2 alone, the full pipeline improves CIDEr-D by 18.2 percentage points on AudioCaps and 11.0 points on Clotho. Compared with Phase~1 alone, the gains are much larger. These results show that the two phases are complementary. Phase~1 transfers acoustic knowledge from the teacher, while Phase~2 adapts the student for caption generation.

\textbf{Generalization beyond captioning.}
To test whether Phase~1 transfers audio competence rather than caption-specific behaviour, we evaluate binary Clotho-AQA~\cite{Lipping2022ClothoAQAAC} zero-shot (Table~\ref{tab:aqa}); no zero-shot checkpoint sees any question, answer or QA-format supervision. After Phase~1, \emph{CARD}$^{*}$ reaches 76.98\% against 60.47\% without distillation and 80.58\% for the encoder-kept SLAM-AAC, so distillation recovers most of the gap to the encoder-based model. Neither Phase~2 fork changes this significantly (78.64\% for Clotho, 75.10\% for AudioCaps), indicating that Phase~2 adapts output format rather than supplying the capability being measured. Fine-tuning directly on Clotho-AQA reaches 80.72\%, so the zero-shot model already attains 95\% of the benchmark-specialized one.

\begin{table}[!h]
\centering
\caption{Binary Clotho-AQA accuracy (\%) on 1{,}442 yes/no test questions.}
\label{tab:aqa}
\setlength{\tabcolsep}{4pt}
\small
\begin{tabular}{l ccc}
\toprule
\textbf{Training} & \textit{No Distill} & \textit{CARD$^{*}$} & SLAM-AAC \\
\midrule
Phase 1                       & 60.47 & 76.98 & \textbf{80.58} \\
Phase 1 + 2 (AudioCaps)       & 62.90 & 75.10 & \textbf{76.63} \\
Phase 1 + 2 (Clotho)          & 66.71 & \textbf{78.64} & 77.41 \\ %
\midrule
Fine-tuned on Clotho-AQA      & 73.51 & 80.72 & \textbf{83.70} \\
\bottomrule
\end{tabular}
\end{table}

\section{Conclusion}

We presented CARD, an encoder-free framework for automated audio captioning. CARD transfers acoustic knowledge from a frozen CLAP teacher into a lightweight audio projector and a LoRA-adapted LLM during training. Our results show that knowledge distillation is important for recovering the performance lost by removing the audio encoder, but its effectiveness depends strongly on where teacher supervision is applied. LLM-only distillation gives negligible gains over the non-distilled encoder-free model, while projector-only distillation with the full teacher hierarchy degrades performance. In contrast, supervising the projector with early teacher stages substantially improves performance, showing that the projector benefits most from lower-level acoustic representations.
The full CARD model combines early-stage projector supervision with later-stage language-model supervision and achieves the best performance among all encoder-free variants. It improves AudioCaps CIDEr-D from 43.2\% to 55.4\% and Clotho CIDEr-D from 22.3\% to 27.5\%, with consistent gains across other captioning metrics. These results suggest that projector-level and language-model-level distillation provide complementary supervision. Overall, CARD demonstrates a promising approach to encoder-free audio captioning by transferring perceptual and semantic audio knowledge into the student model during training, removing the need for a separate audio encoder at inference.

\section*{Acknowledgments}
This work was supported by a Knowledge Transfer Partnership project (Project No. 10131983), funded by UKRI through Innovate UK in collaboration with Hivedome. This work was also partially funded by UKRI and the European Union’s Horizon Europe Research and Innovation Programme through the ELOQUENCE project (Grant Agreement No. 101135916).

\section*{AI-Generated Content Disclosure}

The authors used generative AI tools solely for grammar correction and language refinement. No AI-generated content was used for the development of research ideas, analysis, interpretation of results, or scientific conclusions. The authors reviewed and approved all revised text and take full responsibility for the content of the manuscript.

\bibliographystyle{IEEEtran}
\bibliography{mybib}

\end{document}